\documentclass[aps,prapplied,
  superscriptaddress,
  twocolumn,
  floatfix
]{revtex4-2}

\usepackage[utf8]{inputenc}
\usepackage[english]{babel}
\usepackage[caption=false]{subfig}
\usepackage{upgreek,stix,graphicx,amsmath,amssymb,dcolumn,contour,siunitx,chemmacros,xspace,comment,physics,lipsum}
\usepackage{todonotes}
\usepackage[hidelinks]{hyperref}
\usepackage[normalem]{ulem}

\renewcommand{\selectlanguage}[1]{}

\setuptodonotes{inline,color=tabgreen!50}

\usepackage{etoolbox}
\robustify{\subref}

\graphicspath{{./}{./img/}{./img/tikz/}}

\newcommand\Figref[1]{Figure~\ref{#1}}   
\newcommand\figref[1]{Fig.~\ref{#1}}     
\newcommand\eqnref[1]{Eq.~\ref{#1}}      
\newcommand\tabref[1]{Table~\ref{#1}}    
\newcommand\appref[1]{Appendix~\ref{#1}} 

\definecolor{tabblue}{HTML}{1f77b4}
\definecolor{taborange}{HTML}{ff7f0e}
\definecolor{tabred}{HTML}{d62728}
\definecolor{tabgreen}{HTML}{2ca02c}
\definecolor{tabcyan}{HTML}{17becf}
\definecolor{tabolive}{HTML}{bcbd22}
\definecolor{tabpurple}{HTML}{9467bd}

\newcommand{\approxprop}{\mathrel{\vcenter{
  \offinterlineskip\halign{\hfil$##$\cr
    \propto\cr\noalign{\kern2pt}\sim\cr\noalign{\kern-2pt}}}}}

\newcommand\emx[1]{\ensuremath{#1}\xspace}

\newcommand\timevar{\emx{t}}
\newcommand\freq{\emx{f}}
\newcommand\freqeig{\emx{f_0}}

\newcommand\omeig{\emx{\omega_0}}
\newcommand\omdrive{\emx{\omega_d}}
\newcommand\amp{\emx{A}}
\newcommand\ampdrive{\emx{A_d}}
\newcommand\damping{\emx{\Gamma}}
\newcommand\dampingO{\emx{\Gamma_0}}

\newcommand\fluence{\emx{\Phi}}
\newcommand\fluenceOstress{\emx{\Phi_{0,\stress}}}
\newcommand\fluenceOdamping{\emx{\Phi_{0,\damping}}}
\newcommand\stress{\emx{\sigma}}
\newcommand\stressO{\emx{\sigma_0}}
\newcommand\modeno{\emx{n}}
\newcommand\youngs{\emx{E}}
\newcommand\thick{\emx{h}}
\newcommand\length{\emx{l}}
\newcommand\dens{\emx{\rho}}

\newcommand\Vc{\emx{\mathrm{V}_\mathrm{C}}}
\newcommand\Vsi{\emx{\mathrm{V}_\mathrm{Si}}}
\newcommand\VV{\emx{\mathrm{VV}}}

\newcommand\figscale{0.79}

\begin{document}


\title{Effect of Helium Ion Implantation on 3C-SiC Nanomechanical String Resonators}

\author{Philipp Bredol}
\email{philipp.bredol@tum.de}
\affiliation{Department of Electrical Engineering, 
             School of Computation, Information and Technology,
             Technical University of Munich,
             85748 Garching, Germany}
\author{Felix David}
\affiliation{Department of Electrical Engineering, 
             School of Computation, Information and Technology,
             Technical University of Munich,
             85748 Garching, Germany}
\author{Nagesh S. Jagtap}
\affiliation{Institute of Ion Beam Physics and Materials Research,
             Helmholtz-Zentrum Dresden - Rossendorf,
             01328 Dresden, Germany}
\author{Yannick S. Kla\ss}
\affiliation{Department of Electrical Engineering, 
             School of Computation, Information and Technology,
             Technical University of Munich,
             85748 Garching, Germany}
\author{Georgy V. Astakhov}
\email{g.astakhov@hzdr.de}
\affiliation{Institute of Ion Beam Physics and Materials Research,
             Helmholtz-Zentrum Dresden - Rossendorf,
             01328 Dresden, Germany}
\author{Artur Erbe}
\email{a.erbe@hzdr.de }
\affiliation{Institute of Ion Beam Physics and Materials Research,
             Helmholtz-Zentrum Dresden - Rossendorf,
             01328 Dresden, Germany}
\author{Eva M. Weig}
\email{eva.weig@tum.de}
\affiliation{Department of Electrical Engineering, 
             School of Computation, Information and Technology,
             Technical University of Munich,
             85748 Garching, Germany}
\affiliation{Munich Center for Quantum Science and 
             Technology (MCQST), 80799 Munich, Germany}
\affiliation{TUM Center for Quantum Engineering (ZQE), 
             85748 Garching, Germany}

\date{\today}


\begin{abstract}
Hybrid quantum devices enable novel functionalities by combining the benefits of different subsystems. Particularly, point defects in nanomechanical resonators made of diamond or silicon carbide (SiC) have been proposed for precise magnetic field sensing and as versatile quantum transducers.
However, the realization of a hybrid system may involve tradeoffs in the performance of the constituent subsystems. In a spin-mechanical system, the mechanical properties of the resonator may suffer from the presence of engineered defects in the crystal lattice. This may severely restrict the performance of the resulting device and needs to be carefully explored.
Here, we focus on the impact of defects on high Q nanomechanical string resonators made of pre-stressed 3C-SiC grown on Si(111). We use helium ion implantation to create point defects and study their accumulated effect on the mechanical performance. Using Euler-Bernoulli beam theory, we present a method to determine Young's modulus and the pre-stress of the strings.
We find that Young's modulus is not modified by implantation. Under implantation doses relevant for single defect or defect ensemble generation, both tensile stress and damping rate also remain unaltered. For higher implantation dose, both exhibit a characteristic change.
\end{abstract}

\maketitle

\section{Introduction}

Using quantum phenomena such as coherence, superposition, interference, and entanglement, today's quantum technology can create, trap, manipulate, and detect single particles such as photons, phonons, and spins \cite{acin_quantum_2018,van_deventer_towards_2022}. Hybrid quantum devices combine different subsystems to go beyond the limitations that their components face in stand-alone applications and have been theoretically and experimentally investigated \cite{kurizki_quantum_2015,coupl-spins-to-nanom}. Current nanofabrication techniques enable the coupling of mechanical resonators to electromagnetic radiation by integrating mechanical resonators in optical cavities and superconducting microwave circuits, making cavity optomechanical systems \cite{cavit-optom} a potential candidate for future quantum technologies. 
%
Further, hybrid spin-mechanical systems (HSMS) enable the coupling of phonons from the mechanical mode of a membrane or cantilever to the spin, e.g.\ of a point defect. HSMS with diamond using nitrogen-vacancy (NV) centers have been studied
\cite{topic-revie-spins,coupl-spins-to-nanom}. However, challenges persist in the growth, fabrication, and device processing of diamond \cite{mater-chall-for}.

On the other hand, SiC is a technologically and industrially established material in the field of power electronics. More importantly, several polytypes of SiC have gained importance as a material for nanoelectromechanical systems (NEMS) \cite{yang_monocrystalline_2001,zorman_silicon_2002,yang_zeptogram_scale_2006,li_ultra_sensitive_2007,
micro-with-qfact,engin-the-dissi-of,deter-young-modul}. Also, it hosts point defects with highly coherent spin degrees-of-freedom such as the silicon vacancy (\Vsi) \cite{10.1038/nphys2826, nagy_high_fidelity_2019} and the divacancy (\VV) \cite{falk_polytype_2013}. Such point defects can be created efficiently by ion implantation techniques with nanometer precision \cite{10.1021/acs.nanolett.6b05395, he_maskless_2022}.

Our long-term goal is to realize a hybrid spin-mechanical system to improve magnetic field sensing, e.g. based on spins associated with \Vsi in a nanomechanical resonator made of 4H-SiC \cite{optic-detec-spinm}. To  obtain a thorough understanding of the impact of defect generation on the mechanical performance, we use nanoresonators fabricated in 3C-SiC grown on Si \cite{yang_monocrystalline_2001,zorman_silicon_2002,yang_zeptogram_scale_2006,li_ultra_sensitive_2007}. This material platform provides well-established and reliable process routines for freely-suspended nanostructures. In addition, 3C-SiC grown on Si(111) features a strong tensile pre-stress owing to a \SI{20}{\percent} lattice mismatch \cite{molec-beam-epita}. Pre-stress may enhance the mechanical quality factor by orders of magnitude due to dissipation dilution \cite{brown-motio-of-a,dampi-of-nanom-reson,contr-of-mater-dampi}, which was first reported in amorphous \ch{SiN} nanomechanical resonators \cite{high-quali-facto} and later extended to crystalline \ch{InGaP} \cite{tensi-inxga-membr},
\ch{Si} \cite{strai-cryst-nanom},
\ch{AlN} \cite{nanom-cryst-aln}, 
\ch{SiC} \cite{micro-with-qfact,engin-the-dissi-of,deter-young-modul},
and amorphous \ch{SiC} \cite{highs-amorp-silic}.
Thus, pre-stressed 3C-SiC nanomechanical string resonators serve as an ideal platform for detecting small changes in mechanical properties with high precision while our findings can be qualitatively applied to other polytypes of SiC.

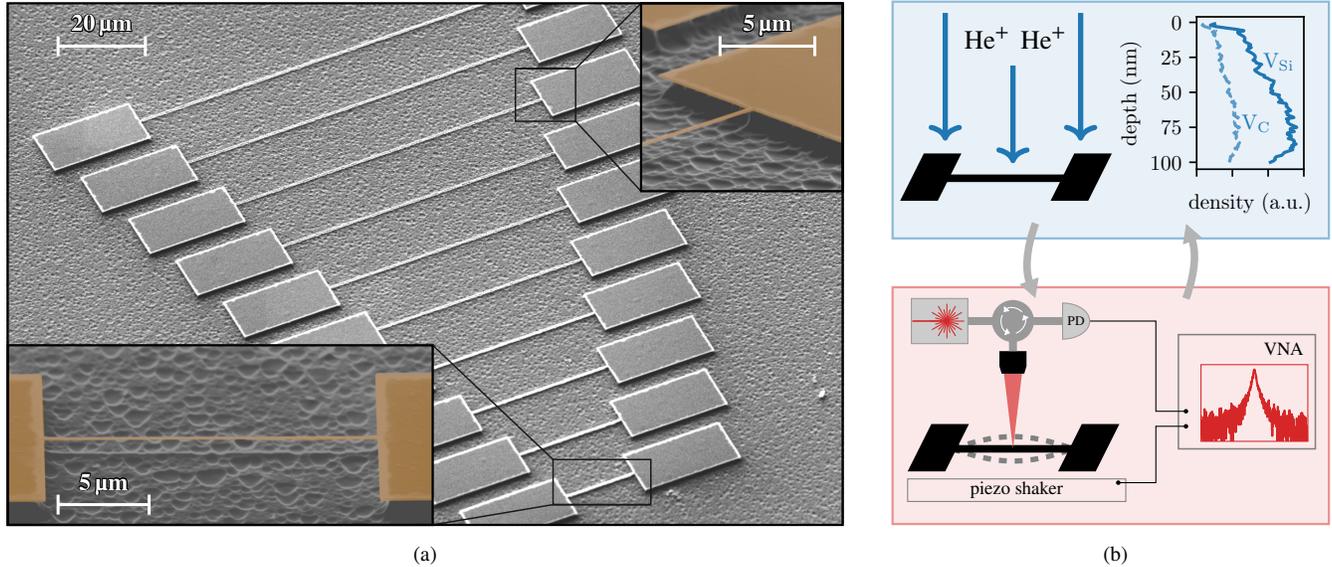
\begin{figure*}
  \centering
  \subfloat[][]{\begin{tikzpicture}
    \node at (0,0) {\includegraphics[width=.62\linewidth]{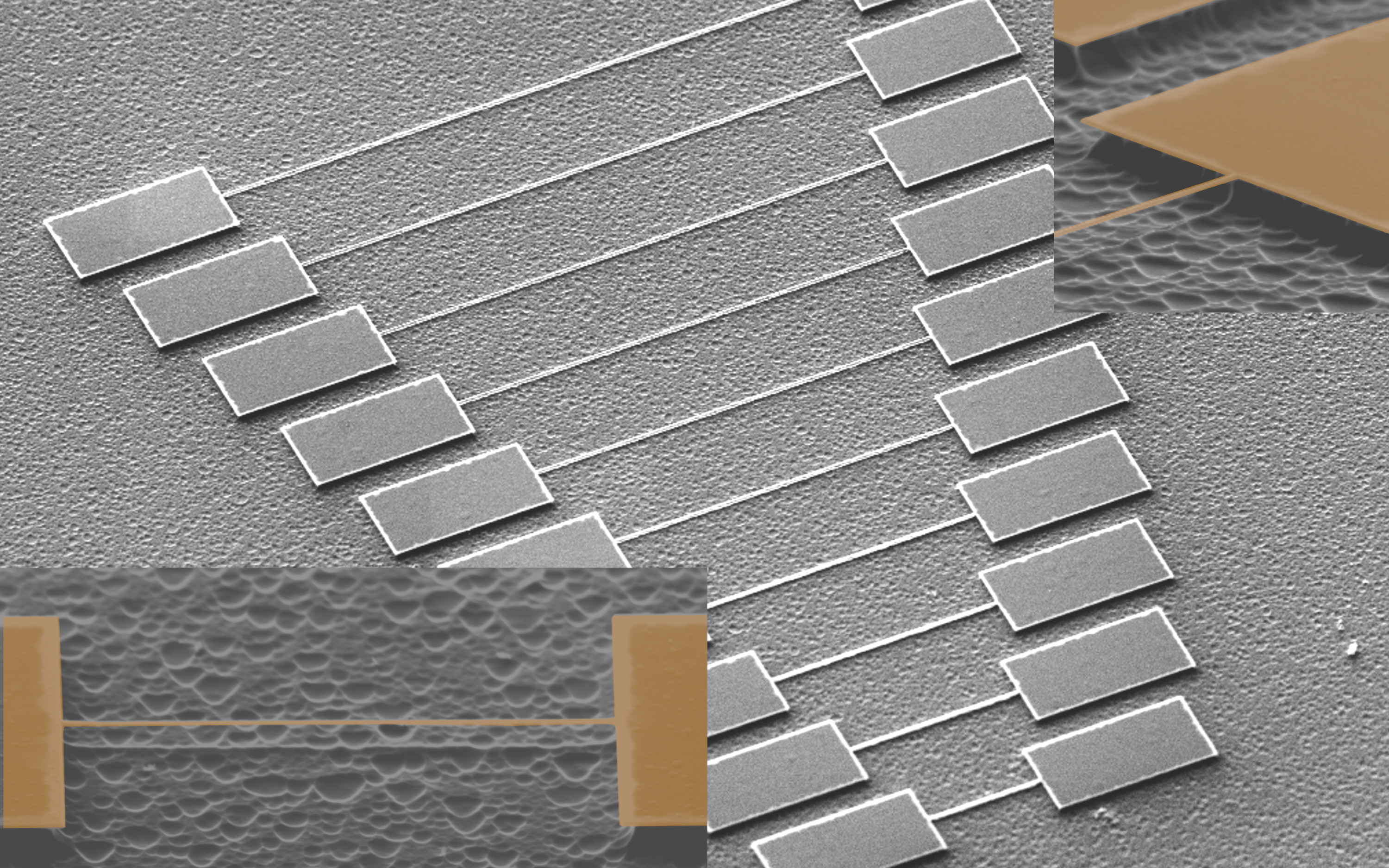}};
    \draw[|-|,line width=.4mm,white] (-4.9,2.9) to node[white,anchor=south]
                      {\contour{black}{\textbf{20\,\textmu m}}} (-3.7,2.9);
    \draw[|-|,line width=.4mm,white] (-4.9,-3.2) to node[white,anchor=south]
                      {\contour{black}{\textbf{5\,\textmu m}}} (-3.65,-3.2);
    \draw[|-|,line width=.4mm,white] (3.9,2.9) to node[white,anchor=south]
                      {\contour{black}{\textbf{5\,\textmu m}}} (5.2,2.9);
    \draw[line width=.2mm,black!90] (.1,-1.08) -- (1.7,-2.6) 
                  rectangle (3,-3.2) -- (1.7,-3.2) -- (.1,-3.47);
    \draw[line width=.2mm,black!90] (2.87,.97) -- (2,1.9) 
                  rectangle (1.2,2.6) -- (2,2.6) -- (2.87,3.47);
    \draw[line width=.3mm] (-5.55,-3.47) rectangle (5.55,3.47);
    \draw[line width=.3mm] (2.87,.97) rectangle (5.55,3.47);
    \draw[line width=.3mm] (-5.55,-3.47) rectangle (.1,-1.08);
    \end{tikzpicture}\label{subfig:sample}}
  \hfill{}
  \subfloat[][]{\begin{tikzpicture}
    \node at (0,0) {\includegraphics[scale=1]{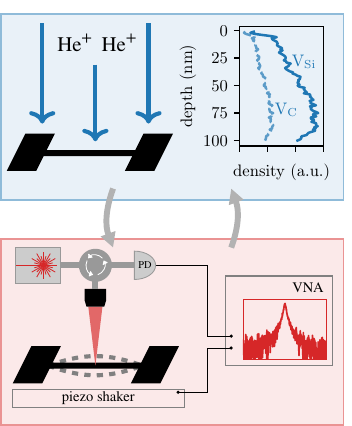}};
    \end{tikzpicture}\label{subfig:measure-implant}}
  \caption{
      \subref{subfig:sample}~Scanning electron micrograph of an array of string resonators with lengths between \SI{20}{\um} and \SI{110}{\um}. Insets show false color tilted-view close-ups of individual strings. Strings and clamping pads are highlighted in orange, the ``shadow'' underneath results from under etching.
      \subref{subfig:measure-implant}~Schematic illustration of the consecutive measurement procedure alternating between He ion implantation (top) and nanomechanical characterization by means of piezo actuation and optical interferometric detection (bottom). The upper right inset shows simulated depth profiles of the normalized density of silicon (\Vsi) and carbon vacancies (\Vc) created by the implantation. 
  }
  \label{fig:sample-setup}
\end{figure*}

To systematically evaluate trends in the mechanical performance, we chose an iterative approach that allows to increase the defect concentration in the resonators under investigation. We start by fabricating nanomechanical string resonators and carefully characterize their vibrational properties. A comprehensive eigenmode analysis is performed to extract all nanomechanical figures of merit, namely the tensile stress \stress, Young's modulus \youngs and damping rate \damping of the individual nanoresonators. Subsequently, we apply broad-beam helium (He) ion implantation (\figref{fig:sample-setup}) to create defects in the nanostring resonators and repeat the mechanical characterization. This procedure is repeated, starting from a low ion fluence with continuously increasing fluence until a clear damage of the nanostring resonators is observed.
Helium is chosen due to its narrow interaction volume of a few ${(\SI{100}{\nm})^3}$ and minimal spread of defect formation \cite{allen_review_2021,hlawacek_helium_2016}. This results in the generation of ensembles of \Vsi and carbon vacancies (\Vc), as well as further, less important defects.

Previous research about the effect of implantation induced defects on the mechanical properties was done in the context of nuclear reactor materials and covers the huge fluences that nuclear reactor walls have to withstand \cite{the-role-of-point,effec-of-he-preim,the-influ-of-irrad,accum-and-recov-of}. However, little is known about the low fluences that are applied to create defects for quantum technologies. Our work fills this gap by reporting the effect of helium ion implantation in the low-damage regime on the properties of high-Q nanomechanical resonators.

\section{Methods}

\subsection{Fabrication of Pre-Stressed Nanomechanical String Resonators}

Nanomechanical resonators are fabricated from a 110 nm 3C-SiC (111) thin film that has been epitaxially grown on a Si (111) wafer by NovaSIC. We use electron beam lithography and a combination of dry etching techniques to realize arrays of freely-suspended nanostrings, i.e.\ one-dimensional doubly-clamped mechanical resonators, of different lengths as shown in \figref{fig:sample-setup}.

Importantly, the resonators fabricated feature a strong static tensile pre-stress inherited from the thin film wafer. This pre-stress is created during the epitaxial growth process because of the large mismatch of lattice constants (\SI{20}{\percent}) and thermal expansion coefficients (\SI{8}{\percent}) of the 3C-\ch{SiC} film and the \ch{Si} substrate \cite{molec-beam-epita}. During the further processing steps this strong pre-stress may relax or enhance depending on the device geometry \cite{unive-lengt-depen-of}. For the string resonator devices discussed here, the pre-stress after fabrication lies between \SI{0.75}{\GPa} and \SI{0.85}{\GPa}. We determine the resulting tensile pre-stress of our string resonators, along with the other mechanical figures of merit, by analyzing the mechanical response spectra, as described at the end of this section. For further details see \appref{app:fab}.

\subsection{He Ion Implantation}

Controlled generation of ensembles of \Vsi in the pre-characterized nanostring resonators is accomplished by He broad-beam ion implantation using the DANFYS 1090-50 implanter (\appref{app:implanter}). We choose an implantation energy of \SI{14}{\keV} which corresponds to \SI{95}{\nm} projected range. \Figref{subfig:measure-implant} depicts the normalized densities of created silicon (\Vsi) and carbon vacancies (\Vc) as a function of depth found by a Stopping and Range of Ions in Matter (SRIM) simulation.

The samples are consecutively exposed to He fluences between ${\SI[parse-numbers=false]{10^{12}}{\per\square\cm}}$ and ${\SI[parse-numbers=false]{10^{14}}{\per\square\cm}}$.
According to the SRIM simulation a fluence of \SI[parse-numbers=false]{10^{14}}{\per\square\cm} creates ${\approx\SI{4e20}{\Vsi\per\cubic\cm}}$ and ${\approx\SI{2e20}{\Vc\per\cubic\cm}}$. The actual defect densities obtained are lower because a large number of the created defects immediately anneal again at room temperature \cite{amorp-and-defec,accum-and-recov-of}.

\subsection{Measurement of Mechanical Response Spectra}

\begin{figure*}
  \begin{minipage}[b]{.48\linewidth}
    ~~
    \subfloat[]{\includegraphics[scale=\figscale]{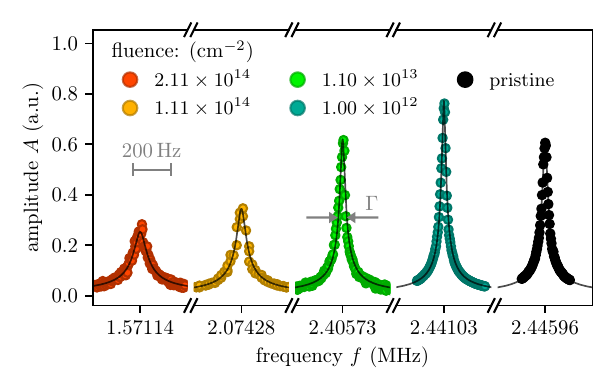}
              \label{subfig:analysis-lorentzians}}\\
    \subfloat[]{\includegraphics[scale=\figscale]{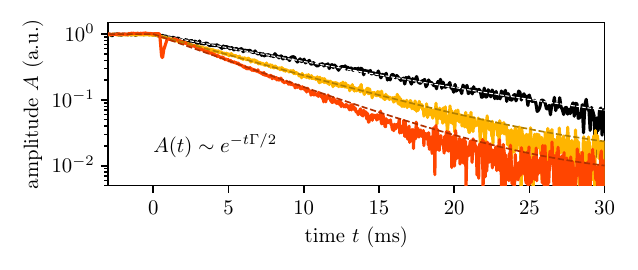}
              \label{subfig:analysis-ringdowns}}
  \end{minipage}
  \begin{minipage}[b]{.48\linewidth}
    \subfloat[]{\includegraphics[scale=\figscale]{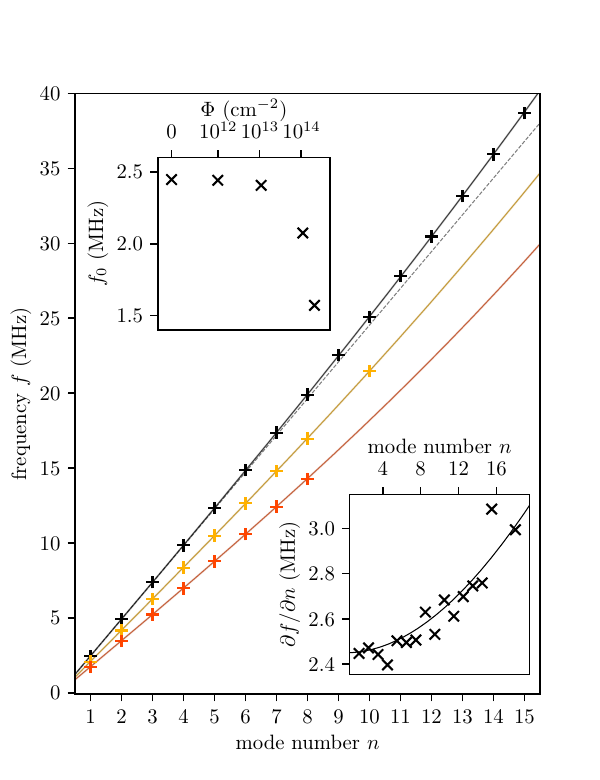}
              \label{subfig:analysis-ebb-fits}}
  \end{minipage}
  \caption{
      \subref{subfig:analysis-lorentzians}~Measured resonance curves (dots) and fits (solid lines) of the fundamental mode of a string with ${\length=\SI{100}{\um}}$ after being subjected to multiple ion irradiation iterations with increasing fluence. The damping rates obtained from the linewidths are ${\damping=\SI{408}{\Hz}}$, \SI{290}{\Hz}, \SI{166}{\Hz}, \SI{132}{\Hz} and \SI{163}{\Hz}, respectively (left to right). All peaks are normalized to enclose the same area ${\int\amp\dd{f}}$.
      \subref{subfig:analysis-ringdowns}~Corresponding ringdown measurements of the same string for three different fluences (color code). The damping rates obtained by fitting an exponential decay curve (dashed lines) are ${\damping=\SI{378}{\Hz}}$, \SI{297}{\Hz} and \SI{199}{\Hz} (bottom to top), agreeing with the damping rates determined from the response curve fits. See \appref{app:conventions} for the used fit functions and conventions.
      \subref{subfig:analysis-ebb-fits}~Resonance frequencies (crosses) of the same string as a function of mode number for three different fluences (color code) and fits to \eqnref{eqn:ebb} (lines). The dashed line follows the linear ${\freq(\modeno)=\modeno\freqeig}$ scaling law. The lower right inset shows the derivative of the resonance frequency with respect to the mode number for the pristine data and fit. Dashed line and inset highlight the small nonlinear component of the measured curves accounting for the bending contribution, which encodes the material's Young's modulus. The upper left inset shows the resonance frequencies from \subref{subfig:analysis-lorentzians} as a function of accumulated fluence.
  }
  \label{fig:data-analysis}
\end{figure*}

To characterize the nanoresonators, we obtain frequency response spectra of a large number of harmonic eigenmodes of all nanostrings. To this end, the drive frequency applied to a piezo shaker underneath the sample is swept across all resonances, and the mechanical response at the drive frequency is read out by means of optical interferometry  via a fast photodetector (PD) with a vector network analyzer (VNA) (\figref{subfig:measure-implant}). The sample and drive piezo are mounted on an xyz-positioner stage to address individual string resonators. To eliminate the effect of gas damping, the measurements are done at pressures below ${\SI[parse-numbers=false]{10^{-3}}{\milli\bar}}$. 
The frequency response of the fundamental out-of-plane mode of an exemplary string with ${\length=\SI{100}{\um}}$ is displayed in \figref{subfig:analysis-lorentzians} for a range of implantation fluences. The upper left inset in \figref{subfig:analysis-ebb-fits} shows these frequencies plotted against the accumulated fluence.
To ensure that the linewidth of the response curve is not broadened by spectral diffusion, we spot-check compatibility with ringdown measurements (\figref{subfig:analysis-ringdowns}).

\subsection{Euler-Bernoulli beam theory fits}

The recorded frequency response curves are fit to \eqnref{eqn:response} to determine the resonance frequency and damping rate as illustrated in \figref{subfig:analysis-lorentzians} (see \appref{app:conventions} for details). Furthermore, we deduce the tensile stress \stress and the Young's modulus \youngs by fitting the resonance frequencies before and after each implantation step to the expected frequencies of an Euler-Bernoulli beam with simply supported boundary conditions,
\begin{align}
  \freq(\modeno)=\frac{\modeno^2\pi}{2\length^2}\sqrt{\frac{\youngs\thick^2}{12\dens}}\sqrt{1+\frac{12\stress\length^2}{\modeno^2\pi^2\youngs\thick^2}},
  \label{eqn:ebb}
\end{align}
with mode number \modeno, string length \length, material mass density \dens, Young's modulus \youngs, tensile pre-stress \stress and string thickness \thick measured along the oscillation direction \cite{funda-of-nanom-reson}. The quantities \stress and \youngs are free fit parameters and \length, \dens and \thick are taken from \tabref{tab:params}. We find that the measured frequencies and the fit function are in agreement (\figref{subfig:analysis-ebb-fits}), thereby justifying the choice of boundary conditions for our devices.

\begin{table}
  \begin{ruledtabular}
  \begin{tabular}{rrrlc}
    & mass density: & \dens~~= & \SI{3.2(1)e3}{\kg\per\cubic\meter} & \cite{henisch2013silicon,harris1995properties} \\
    & beam thickness: & \thick~~= & \SI{110(2)}{\nm} & analogous to \cite{deter-young-modul} \\
    & string length: & \length~~= & ${\length_0\pm\SI{500}{\nm}}$ & analogous to \cite{deter-young-modul} \\
  \end{tabular}
  \end{ruledtabular}
  \caption{Material and sample parameters used for the fits to Euler-Bernoulli beam theory (\eqnref{eqn:ebb}). ${\length_0}$ denotes the nominal length of the lithography design.}
  \label{tab:params}
\end{table}

We estimate the uncertainties of the optimal fit parameter results using Monte-Carlo error propagation. For that purpose, each frequency is re-measured multiple times during the characterization. For each Monte-Carlo run one of these measured frequencies is picked at random for each mode. \length, \dens and \thick are drawn from a normal distribution for each run with the standard deviation given by the respective parameter uncertainty (\tabref{tab:params}). We run the fit for many of these random choices of parameters and frequencies. The uncertainties of \stress and \youngs are then given by the standard deviation of the fit results of all runs.
%
This error propagation method is closely related to the method presented in \cite{deter-young-modul} but more forgiving with respect to outlier frequencies, which helps analyzing the large datasets generated in this work.

Inserting the material parameters of our samples in \eqnref{eqn:ebb}, one finds that the frequencies are ${\freq(\modeno)\approxprop\sqrt{\stress}\modeno}$, i.e.\ the stress is related to the slope of ${\freq(\modeno)}$, which can be fit reliably even with few data points. The Young's modulus \youngs, however, corresponds to a tiny nonlinear component of ${\freq(\modeno)}$, which accounts for the contribution of the bending rigidity to the string's dynamics and becomes apparent only for higher modes (\figref{subfig:analysis-ebb-fits}). This leads to \youngs being more sensitive to the uncertainties of measured frequencies, geometry and material parameters than the fit parameter \stress. Thus, the results for the Young's modulus \youngs tend to show larger relative uncertainties than the results for the stress \stress.

\section{Results and Discussion}

Two samples were fabricated and are referred to as ``sample A'' and ``sample B'' in this manuscript. All visible mechanical modes of 14 different strings were characterized on sample A for the pristine state and at 4 different ion fluences. To evaluate the sample to sample variation of the observed trends, we additionally characterized 15 strings on sample B at two different fluences. The shift of the fundamental mode frequency and the resonance broadening is clearly observed with increasing He implantation fluence \fluence (\figref{subfig:analysis-lorentzians}).

\subsection{Ion-Induced Stress Relaxation}

\begin{figure}
  \centering
  \includegraphics[scale=\figscale]{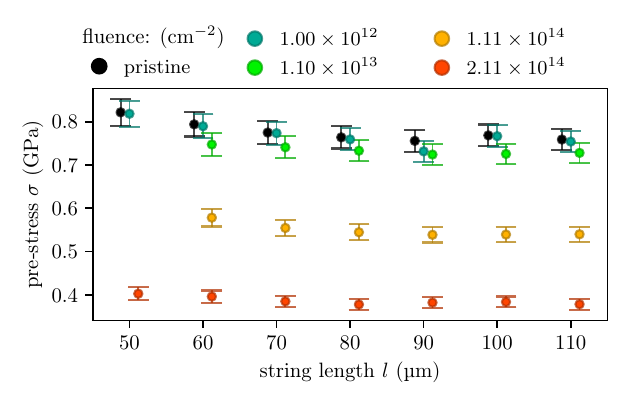}
  \caption{
    Tensile pre-stress extracted from each string's frequency response (see \figref{fig:data-analysis}) as a function of string length for sample A. The color indicates the accumulated fluence, the error bars show the combined uncertainty of included geometry parameters, material parameters and fit uncertainty. Data is slightly $x$-shifted for clarity.
  }
  \label{fig:effect-on-stress-vs-length}
\end{figure}

\begin{figure}
  \subfloat[]{\includegraphics[scale=\figscale]{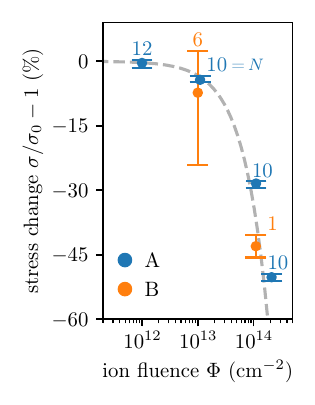}
             \label{subfig:effect-on-stress}}
  \subfloat[]{\includegraphics[scale=\figscale]{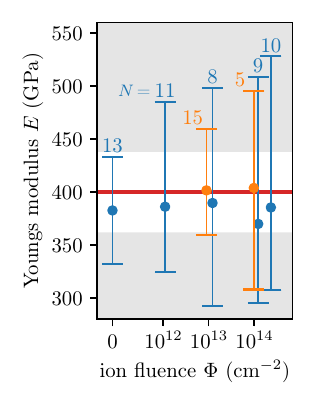}
             \label{subfig:effect-on-youngs-mod}}
  \caption{
    \subref{subfig:effect-on-stress}~Tensile pre-stress \stress as a function of accumulated fluence \fluence normalized to the initial pre-stress \stressO of the pristine string resonator averaged over string length for samples A (blue) and B (orange). The error bars indicate the standard deviation. The numbers close to the error bars indicate the number of underlying data points. The dashed line traces the relation ${\stress/\stressO=1-\fluence/\fluenceOstress}$ with ${\fluenceOstress=\SI{3e14}{\per\cm\squared}}$.
    \subref{subfig:effect-on-youngs-mod}~Equivalent plot of Young's modulus. The red line and shaded region indicate ${\youngs=\SI{400(38)}{\GPa}}$ as determined in \cite{deter-young-modul}.}
  \label{fig:effect-on-stress-and-youngs-mod}
\end{figure}

\Figref{fig:effect-on-stress-vs-length} shows the stress obtained from the Euler-Bernoulli beam fits (\eqnref{eqn:ebb}) for all investigated fluences. For the pristine sample, the pre-stress increases with shortening string lengths, which is a side-effect of stress redistribution during etching \cite{unive-lengt-depen-of}. After the initial ${\fluence=\SI[parse-numbers=false]{10^{12}}{\per\square\cm}}$ implantation, the pre-stress is still close to the pre-stress of the pristine sample, indicating a negligible shift of mechanical eigenfrequencies up to this fluence. During the further implantation runs, the stress significantly relaxes (\figref{subfig:effect-on-stress}). The dependence of this stress relaxation on the fluence follows the simple phenomenological relation
\begin{align}
  \frac{\stress(\fluence)}{\stressO}=1-\frac{\fluence}{\fluenceOstress},
\end{align}
with \fluence denoting the total He ion fluence that the sample was exposed to, the remaining pre-stress at a given fluence ${\stress(\fluence)}$, the pre-stress  of the pristine sample \stressO and the phenomenological parameter \fluenceOstress, which corresponds to the fluence at which the pre-stress would be fully relaxed (${=\SI{3e14}{\per\cm\squared}}$ for our samples). This relation is compatible with volumetric material swelling in proportion to the implanted He ion fluence, which agrees with previous measurements \cite{the-influ-of-irrad} and simulations \cite{the-role-of-point}.

\subsection{Independence of Young's Modulus}

Whereas we observe a significant relaxation of the pre-stress, Young's modulus stays constant within the error margins for both samples and all employed fluences (\figref{subfig:effect-on-youngs-mod}). This observation agrees with previous experiments determining the Young's modulus of \ch{SiC} by nanoindentation \cite{effec-of-heliu-impla,effec-of-he-preim} and molecular dynamics simulations \cite{the-role-of-point}.
%
The uncertainties of the extracted values for Young's modulus tend to increase with the accumulated ion fluence because less mechanical modes were visible the more as the sample was transferred between labs and setups during the consecutive measurement protocol. 

The number of data points used to calculate the standard deviation in the two plots shown in \figref{fig:effect-on-stress-and-youngs-mod} differs even though the same dataset was used for both. On the one hand, obtaining a meaningful \youngs requires a large number of measurable mechanical modes, whereas \stress can be determined with a few modes already. On the other hand ${\stress/\stressO}$ requires the stress of implanted and pristine state to be known.

It is important to note that the Young's modulus values we measure are based on the bending rigidity of the nanostring resonators. This implicitly assumes a slender string with rectangular cross section made from a material with homogenous elastic properties. These assumptions are perfectly justified for the pristine data. After implantation, however, the latter assumption is not obviously fulfilled. We expect stronger material modification close to the substrate-facing surface of the nanostring resonators than in the rest of the material, due to the chosen implantation profile (\figref{subfig:measure-implant}). The fact that we observe no significant change in Young's modulus for the investigated fluence range indicates that the material modifications caused negligible changes in the elastic properties, hence the assumption of homogeneous elastic properties is also justified after implantation, allowing us to safely deduce Young's modulus from the bending rigidity.

\subsection{Ion-Induced Mechanical Damping}

Apart from pre-stress and Young's modulus, we study the mechanical damping rates of the nanostring resonators as a function of accumulated fluence. The damping rates shown here are obtained from the linewidth fits (\eqnref{eqn:response}) of the response curves (\figref{subfig:analysis-lorentzians}). We spot-check the consistency of the obtained damping rates with ringdown measurements (\figref{subfig:analysis-ringdowns}), to make sure the measured linewidths are reflecting the mechanical damping rates and are not broadened by spectral diffusion, instrumentation noise, resolution limits, etc.

\begin{figure}
  \includegraphics[scale=\figscale]{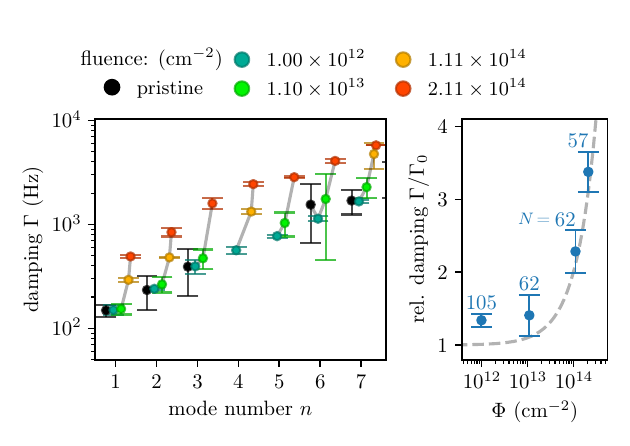}\\[-.6cm]
  \hspace{.18\linewidth}
  \subfloat[]{\label{subfig:damping-vs-modeno}}
  \hspace{.43\linewidth}
  \subfloat[]{\label{subfig:rel-damping-vs-fluence}}
  \caption{
      \subref{subfig:damping-vs-modeno}~Dependence of the mechanical damping rate \damping on the accumulated fluence \fluence for the first 7 flexural out-of-plane modes of an exemplary string with ${\length=\SI{100}{\um}}$.
      \subref{subfig:rel-damping-vs-fluence}~Damping \damping relative to the respective pristine \dampingO as a function of \fluence averaged over all string lengths and mode numbers measured on sample A. The error bars indicate the standard deviation. The numbers close to the error bar indicate the number of underlying measured values of \damping. The dashed line traces the relation ${\damping/\dampingO=1+\fluence/\fluenceOdamping}$ with ${\fluenceOdamping=10^{14}\,\si{\per\square\cm}}$.}
  \label{fig:damping-vs-modeno}
\end{figure}

\Figref{fig:damping-vs-modeno} shows that the damping rates stay constant within experimental errors up to an accumulated fluence of ${\approx\SI[parse-numbers=false]{10^{13}}{\per\square\cm}}$. For higher fluences, however, the damping increases rapidly, indicating that the internal friction caused by the irradiation damage outweighs the other friction mechanisms.

The damping rate of nanostring resonators is sensitive to contamination or other forms of degradation which may occur while mounting, unmounting and transporting the sample between the different setups. This limits the repeatability of the damping rate measurement to \SI{\pm 30}{\percent} in our experiment. The error bars in \figref{subfig:rel-damping-vs-fluence} reflect the standard deviations of measurements done in one run and do not take the repeatability into account. It is worth noting that the two data points of \figref{subfig:rel-damping-vs-fluence} with ${\fluence\leq\SI{1.1e13}{\per\square\cm}}$ seem to disagree with the trend indicated by the dashed line. However, considering the repeatability of \SI{\pm 30}{\percent}, both points are in good agreement with the trend.

Similar to the behavior of the string pre-stress, the relative damping rate ${\damping/\dampingO}$ depends linearly on the accumulated fluence (dashed line in \figref{subfig:rel-damping-vs-fluence})
\begin{align}
  \frac{\damping\qty(\fluence)}{\dampingO} = 1+\frac{\fluence}{\fluenceOdamping},
\end{align}
with \fluence denoting the accumulated fluence, the damping at a given fluence ${\damping\qty(\fluence)}$, the corresponding damping in the pristine state \dampingO and the phenomenological parameter \fluenceOdamping. The slope of this trend is given by ${\fluenceOdamping=\SI[parse-numbers=false]{10^{14}}{\per\square\cm}}$ for the ion induced damping increase and ${\fluenceOstress=\SI{3e14}{\per\square\cm}}$ for the stress relaxation. It is known that the various possible point defects contribute differently to the volumetric swelling of \ch{SiC} \cite{the-role-of-point}. Similarly, their contribution to the inner mechanical friction likely varies, too, explaning the different scaling of stress relaxation and damping increase with accumulated fluence.

The optimum implantation fluence to create the \Vsi ensembles is in the range of ${\SIrange[parse-numbers=false]{5\times10^{13}}{10^{14}}{\per\square\cm}}$ \cite{10.48550/arxiv.2310.18843}. We observe a moderate increase of \damping for this fluence regime. Thus, for hybrid spin-mechanical devices based on defect ensembles additional mechanical damping caused by the implantation is to be expected. 
However, the typical implantation fluences required to create isolated optically active \Vsi are in the range of ${\SI[parse-numbers=false]{10^{11}}{\per\square\cm}}$ \cite{10.1038/s41563-021-01148-3}. This fluence is far below the observed threshold of ${\approx\SI[parse-numbers=false]{10^{13}}{\per\square\cm}}$, at which additional mechanical damping starts to become noticeable. Therefore, for applications based on single defects additional ion-induced damping is expected to not play a dominant role.

\section{Conclusion}

We present measurements of the eigenfrequency and damping rates of the flexural out-of-plane modes of nanomechanical string resonators made of strongly pre-stressed 3C-\ch{SiC} as a function of accumulated helium ion fluence $\fluence$ under broad-beam ion implantation. We obtain the tensile pre-stress and Young's modulus by fitting the measured eigenfrequencies to Euler-Bernoulli beam theory. The presented method can also be applied to other types of mechanical nanodevices, such as cantilevers and membranes.

We find that the pre-stress relaxes for ${\fluence>\SI[parse-numbers=false]{10^{13}}{\per\square\cm}}$ and drops to \SI{50}{\percent} of the original value for the highest investigated fluence ${\fluence=\SI{2.11e14}{\per\square\cm}}$. The stress relaxation agrees with the well-known volumetric swelling of \ch{SiC} upon helium implantation. Young's modulus remains unchanged for all investigated implantation fluences.

The damping rate stays constant up to a threshold of ${\fluence\approx\SI[parse-numbers=false]{10^{13}}{\per\square\cm}}$ and increases rapidly for higher fluences. We conclude that creating point defect ensembles for hybrid spin-mechanical devices using ion implantation does not cause excessive additional mechanical damping. When creating single defects, the additional mechanical damping is negligible.

The observed stress relaxation shows that the resonance frequencies of pre-stressed string resonators are widely tunable using \ch{He} implantation without causing additional damping, which could be used to individually tune resonators by local implantation in a helium ion microscope. Usually, the resonance frequencies of two nominally identical pre-stressed string resonators differ by hundreds of linewidths. Individual tuning in a helium ion microscope would allow to fabricate resonators with resonance frequencies matched better than their linewidths or with well defined frequency splittings to engineer resonant couplings between strings.


\begin{acknowledgements}

We gratefully acknowledge financial support from the Deutsche Forschungsgemeinschaft (DFG, German Research Foundation) through Project Nos. AS 310/9-1, ER 341/17-1, WE 4721/1-1. Support from the Ion Beam Center (IBC) and the Nanofabrication Facilities Rossendorf at IBC of HZDR is gratefully acknowledged for ion implantation and nanofabrication. Support from the Center for Nanotechnology and Nanomaterials (ZNN) Garching is gratefully acknowledged for nanofabrication.

\end{acknowledgements}

\bibliography{main}


\appendix
\section{Fabrication of Nanomechanical String Resonators}
\label{app:fab}

We fabricate pre-stressed nanomechanical string resonators from commercially available 3C-\ch{SiC} thin film wafers grown on \ch{Si}(111) by NovaSiC. The nominal 3C-\ch{SiC} film thickness is \SI{110}{\nm}. We pattern 4 arrays of nanostring resonators as shown in \figref{subfig:sample} on a \mbox{\qtyproduct{5x5}{\mm}} sample using electron-beam lithography and polymethyl methacrylate (PMMA) resist. Each array consists of 10 nanostring resonators with a width of \SI{290}{\nm} (measured on a reference sample) and lengths increasing from \SI{20}{\um} to \SI{110}{\um} in steps of \SI{10}{\um} (\figref{subfig:sample}). We evaporate 30 \si{\nm} of \ch{Cr} followed by a lift-off to obtain the patterned hard mask. Anisotropic reactive ion etching with \ch{SF6} (\SI{2}{sccm}) and argon (\SI{4}{sccm}) at an ICP-Power of \SI{200}{W} and HF-Power of \SI{20}{W} transfers the pattern into the \ch{SiC} thin film. We apply isotropic reactive ion etching to the silicon substrate to release the strings. Therefore, we also use a gas mixture of \ch{SF6} (\SI{30}{sccm}), argon (\SI{5}{sccm}) with HF-Power of \SI{20}{W} and no ICP-Power. We assume a total etching depth of \SI{1.8}{\micro\meter}. Finally, the \ch{Cr} hard-mask is removed using chromium etchant 1020 from Transene.

\section{Ion Implanter DANFYS 1090-50}
\label{app:implanter}

The employed implantation machine is a DANFYS 1090-50 electrostatic air-insulated accelerator produced by Danfysik. The SO140 ion source is integrated directly into the high-voltage terminal. The positively charged ions generated by electron impact ionization in the ion source are accelerated toward the ion beam line. The maximum acceleration voltage that can be achieved is \SI{40}{\kV}. The ion beam is scanned over the sample in horizontal and vertical directions by deflecting plates supplied with a triangle voltage (frequency around 1 kHz) to achieve homogeneous implantation. The vacuum inside the sample chamber is around ${p=\SI{1e-7}{\milli\bar}}$. 

\section{Characterization Setup}
\label{app:setup}

\begin{figure*}
  \includegraphics[width=\linewidth,trim={2cm 13cm 2cm 9cm},clip]{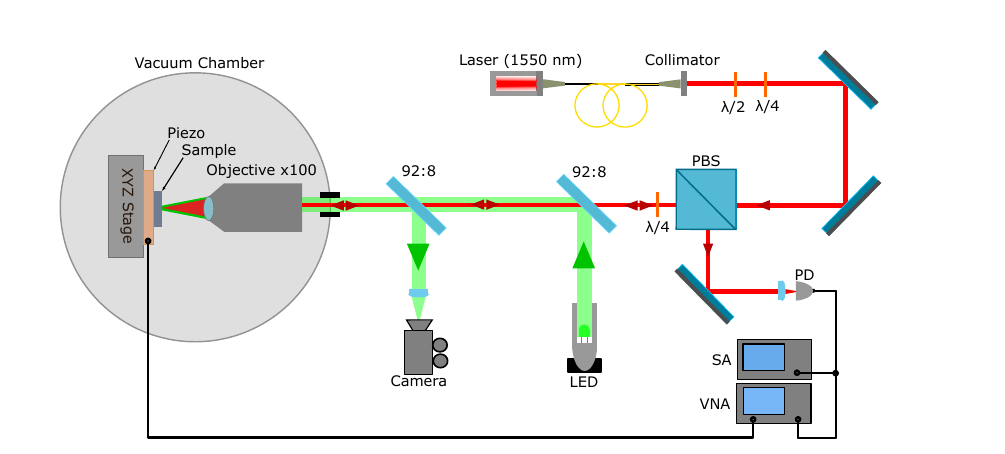}
  \caption{Schematic of the optical measurement setup to measure the deflection of the nanomechanical string resonator. Abbreviations in the sketch: PBS - Polarized Beam Splitter, PD - Photo Detector, SA - Spectrum Analyzer, VNA -  Vector Network Analyzer.}
  \label{fig:appendix-setup}
\end{figure*}

\Figref{fig:appendix-setup} shows a sketch of the interferometric measurement setup. Laser light with \SI{1550}{\nm} sent through half and quarter waveplates to ensure well defined linear polarization, which is fully transmitted by the polarizing beam splitter cube (PBS). The transmitted light is converted to circular polarization by another quarter wave plate and focussed on the sample using a standard microscope objective. 
Since the cross-section of the laser spot is larger than the string's width, only a part of the light is reflected by the string surface, the rest is reflected by the silicon substrate. Due to the interference of these to reflections, vibrations of the string modulate the reflected light intensity. The reflected light is collected by the objective and converted back to linear polarized light by the quarter waveplate, which is fully reflected to the fast photodetector (PD) by the PBS. The electrical signal of the PD is measured by a vector network analyzer for frequency response measurmements or with a spectrum analyzer to conduct ring-down measurements. 
To navigate on the sample and to find the string resonators, an additional LED and a camera are coupled to the beam path with weakly reflecting 92:8 mirrors. The sample is moved by attocube positioners. The high frequency drive is applied by a piezo plate underneath the chip. During the measurement, the focus objective, the sample, and the XYZ-positioner are held in a vacuum better than ${p=\SI{1e-4}{\milli\bar}}$.

\section{Frequency Response and Ringdown Fit Functions}
\label{app:conventions}

We use the conventions implied by the equation of motion for the driven damped harmonic oscillator

\begin{align}
  \ddot{x} + \damping\dot{x} + \omeig^2 x = \ampdrive \cos (\omdrive\timevar),
\end{align}

with the deflection $x$, the damping rate \damping (i.e.\ energy decay rate), the resonance frequency ${\omeig=2\pi\freqeig}$, the driving frequency \omdrive, the driving force normalized to effective mass \ampdrive, and the time \timevar. The fit functions used for the measured amplitude response curves and amplitude ringdown traces are given by

\begin{align}
  \amp_\mathrm{response}\qty(\omdrive) &= \frac{\damping\omeig\amp_0}{ \sqrt{\qty(\omeig^2-\omdrive^2)^2 + \damping^2\omdrive^2} } + \amp_\mathrm{noise},
  \label{eqn:response}\\
  \amp_\mathrm{ringdown}\qty(\timevar) &= \amp_0 e^{-\timevar\damping/2} + \amp_\mathrm{noise},
\end{align}

with the signal amplitude ${\amp_0}$ and the experimental noise floor ${\amp_\mathrm{noise}}$. The parameters ${\amp_0}$, \omeig, \damping and ${\amp_\mathrm{noise}}$ are free fit parameters. All damping rates and amplitudes mentioned in the main text follow these conventions.

\clearpage


\end{document}